%% file: main.tex
\documentclass[11pt,a4paper]{article}
\usepackage[T1]{fontenc}
\usepackage{lmodern}
\usepackage[margin=1in]{geometry}
\usepackage{amsmath,amssymb,amsthm,mathtools}
\numberwithin{equation}{section}
\usepackage{booktabs}
\usepackage{algorithm}
\usepackage{algpseudocode}
\usepackage{microtype}
\usepackage{hyperref}
\hypersetup{colorlinks=true,linkcolor=blue,citecolor=blue,urlcolor=blue,
  pdfauthor={Binh T. Nguyen},
  pdftitle={Kidney Exchange in Deterministic O*(4^(t+o(t))) Time}}
\newtheorem{theorem}{Theorem}
\newtheorem{lemma}[theorem]{Lemma}
\newtheorem{proposition}[theorem]{Proposition}
\newtheorem{corollary}[theorem]{Corollary}
\theoremstyle{definition}
\newtheorem{definition}[theorem]{Definition}

\newcommand{\KEP}{\textsc{Kidney Exchange}}

\newcommand{\len}{\operatorname{len}}
\newcommand{\Query}{\operatorname{Query}}

\title{Kidney Exchange in Deterministic $O^*(4^{t+o(t)})$ Time}
\author{Binh T. Nguyen\\[0.3em]
  \small Department of Computer Science\\
  \small Faculty of Mathematics and Computer Science\\
  \small University of Science, Vietnam National University Ho Chi Minh City\\
  \small Ho Chi Minh City, Vietnam}
\date{November 2026}

\begin{document}
\maketitle

\begin{abstract}
Kidney exchange admits a randomized fixed-parameter algorithm with running
time $O^*(4^t)$, where $t$ is the target number of recipients receiving a
transplant, whereas the best previous deterministic bound is
$O^*(6.855^t)$.  We close this gap at the level of the leading exponential
base by giving a deterministic algorithm running in
\[
    2^{2t+O(\sqrt t\log^2 t)}n^{O(1)}
    = O^*(4^{t+o(t)}).
\]
Our algorithm is built around a dynamic representation of partial exchange
packings.  We show that the semi-feasible packing recurrence underlying the
best previous deterministic algorithm can be executed directly using
Dynamic Representative Sets under a single global cardinality budget $2t$.
The key invariant is a residual extension budget: when a transition inserts
$s$ new vertices, the predecessor has exactly $s$ additional units of
disjoint-extension budget, matching the $s$ vertices that must be excluded
from future use.  This permits the recurrence to be maintained dynamically,
without repeatedly constructing and compressing static representative
families.

A separate obstacle is that a single admissible cycle may already contain
at least $t$ transplants.  We resolve it by giving a deterministic
$O^*(4^{k+o(k)})$ algorithm for computing the minimum size of a simple
directed cycle containing at least $k$ vertices.  This shortest-long-cycle
primitive combines exact-size directed-path optimization with an
optimum-compatible partition argument and keeps the large-cycle
preprocessing within the same exponential base.

Finally, the dynamic representation preserves additive optimization, not
only feasibility.  We obtain a deterministic $O^*(4^{t+o(t)})$ algorithm
for maximizing additive arc utility among feasible packings with exactly
$t$ transplants.  Consequently, if $\tau$ is the maximum achievable number
of transplants, a packing that first maximizes cardinality and then additive
utility can be computed deterministically in
$O^*(4^{\tau+o(\tau)})$ time.
\end{abstract}
\noindent\textbf{Keywords:} kidney exchange, parameterized algorithms, representative sets, directed cycles, fixed-parameter tractability.
\medskip

\input{sections/01_introduction}
\input{sections/02_related_work}
\input{sections/03_preliminaries}
\input{sections/04_slc}
\input{sections/05_semifeasible}
\input{sections/06_drs}
\input{sections/07_main_theorem}
\input{sections/08_conclusion}

\bibliographystyle{plain}
\bibliography{references}

\clearpage
\appendix
\input{appendices/A_slc_reconstruction}
\input{appendices/B_semifeasible_reconstruction}
\input{appendices/C_drs_invariant_details}
\input{appendices/D_weighted_details}
\end{document}

%% file: sections/01_introduction.tex
\section{Introduction}
\label{sec:introduction}

Kidney exchange arranges vertex-disjoint cycles among incompatible donor--recipient pairs and chains initiated by altruistic donors in a directed compatibility graph.  With medical and logistical caps on component lengths, the solution-size parameterization asks whether at least $t$ recipients can receive transplants \cite{ManloveOMalley2014,DickersonEtAl2016,BarkelEtAl2026}.  H\'ebert-Johnson et al. gave a randomized $O^*(4^t)$ algorithm, whereas the best deterministic bound is $O^*(6.855^t)$, due to Tian and Xiao \cite{HebertJohnsonEtAl2024,TianXiao2026}.  We close this gap at the leading exponential base:
\begin{equation}
T_{\KEP}(t,n)=2^{2t+O(\sqrt t\log^2t)}n^{O(1)}=O^*(4^{t+o(t)}).
\end{equation}

Two obstacles drive the proof.  First, bounded cycles and chains interact globally only through vertex-disjointness.  Tian and Xiao encode partial packings by used-vertex families and show that an accepting computation can be witnessed with at most $2t$ used vertices \cite{TianXiao2026}.  We execute their semantic recurrence directly with Nederlof's Dynamic Representative Sets (DRS) \cite{Nederlof2026}.  For a state using $p$ vertices, the required future-disjointness budget is $2t-p$.  Under the single global DRS budget $K=2t$, a transition inserting $s$ vertices satisfies
$
K-(p-s)=(K-p)+s,
$
so a successor query against $Y$ becomes a valid predecessor query against $Y\cup S$.  This residual-budget identity eliminates repeated static representative-family compression.

Second, before the bounded-component DP applies, one admissible component may already contain $t$ transplants.  Chains can be truncated, but cycles cannot: with an upper cycle cap, one needs the minimum size
\begin{equation}
\lambda_k(G)=\min\{|V(C)|: C\text{ simple directed cycle},\ |V(C)|\ge k\}.
\end{equation}
We compute $\lambda_k(G)$ deterministically in $O^*(4^{k+o(k)})$ time by combining exact-size directed-path optimization with Zehavi's Long Directed Cycle partition framework \cite{Zehavi2016,Nederlof2026}.  This keeps large-component preprocessing within the target base.

The dynamic representation also supports min-plus optimization.  With additive arc utilities, Exact-$t$ Weighted Kidney Exchange is solvable deterministically in $O^*(4^{t+o(t)})$ time.  Consequently, if $\tau$ is the maximum achievable transplant cardinality, cardinality-first additive-utility optimization takes $O^*(4^{\tau+o(\tau)})$ time.

\paragraph{Contributions.}
Our contributions are: (i) the deterministic $O^*(4^{t+o(t)})$ KEP algorithm; (ii) the residual-budget invariant realizing the semi-feasible recurrence under one DRS budget $K=2t$; (iii) to the best of our knowledge, the first deterministic $O^*(4^{k+o(k)})$ algorithm for computing the minimum size of a directed cycle above a prescribed threshold; and (iv) the exact-cardinality and cardinality-first additive-utility extensions.

\paragraph{Context and organization.}
Earlier parameterized work progressed from FPT algorithms \cite{MaitiDey2022} to randomized $O^*(4^t)$ and deterministic $O^*(14.34^t)$ bounds \cite{HebertJohnsonEtAl2024}, followed by the representative-set approach of Tian and Xiao \cite{TianXiao2026}; Banik et al. give a related color-coding approach \cite{BanikEtAl2026}.  Long directed paths and cycles have likewise been studied via single-exponential, representative-family, and partition methods \cite{GabowNie2008,FominEtAl2016,Zehavi2016,FominEtAl2018LongPath}.  Section~\ref{sec:related} gives the closest comparisons; Sections~\ref{sec:slc}--\ref{sec:main} give the algorithm and proofs.  Detailed reconstructions appear in Appendices~A--D.

%% file: sections/02_related_work.tex
\section{Related Work}
\label{sec:related}

\paragraph{Parameterized Kidney Exchange.}
Maiti and Dey established fixed-parameter tractability in the number of transplants \cite{MaitiDey2022}.  H\'ebert-Johnson et al. obtained randomized $O^*(4^t)$ and deterministic $O^*(14.34^t)$ algorithms \cite{HebertJohnsonEtAl2024}.  Tian and Xiao subsequently obtained the current deterministic $O^*(6.855^t)$ bound using a semi-feasible packing DP and static representative families; their analysis also supplies the bounded regime $t\le c(\mathcal D)\le2t$ and the refined cardinality index $p\le2t$ used here \cite{TianXiao2026}.  Banik et al. proposed a related deterministic color-coding approach; Tian and Xiao identify a witness-size issue in that analysis \cite{BanikEtAl2026,TianXiao2026}.  Our contribution is not a new witness bound, but a dynamic realization of the semantic recurrence at global budget $K=2t$.

\paragraph{Representative sets and long cycles.}
Representative-family methods are a standard tool in parameterized algorithms \cite{FominEtAl2016}.  Nederlof's Dynamic Representative Sets support union, element convolution, and bounded disjoint-extension queries in $2^{K+O(\sqrt K\log^2K)}$ dependence on the global budget, and extend to additively idempotent semirings \cite{Nederlof2026}.  We exploit the exact alignment between this query budget and the $2t-p$ future-extension budget of the KEP states.

For Long Directed Cycle, representative-family methods give deterministic single-exponential bounds \cite{FominEtAl2016}, while Zehavi's partition reduction separates an exact-size regime from a difficult long-cycle regime \cite{Zehavi2016}; Fomin et al. also study minimum long $(s,t)$-paths and obtain a deterministic $O^*(4.884^k)$ decision bound \cite{FominEtAl2018LongPath}.  Zehavi together with Nederlof yields a $4^{k+o(k)}$ decision route.  Our shortest-long-cycle result recovers the \emph{minimum} feasible cycle size at this deterministic base, which is needed when KEP also imposes an upper cycle cap.

\paragraph{Weighted Kidney Exchange.}
Maximum-weight and hierarchical objectives are well established in kidney exchange, both in cycle-packing models and exact optimization systems \cite{BiroManloveRizzi2009,LinEtAl2019,DelormeEtAl2024}.  We do not introduce additive utility as an objective; rather, we show that exact-$t$ additive optimization, and hence cardinality-first additive utility, inherits the same deterministic $4^{t+o(t)}$ parameter dependence as our feasibility algorithm.

%% file: sections/03_preliminaries.tex
\section{Preliminaries}
\label{sec:prelim}

Let \(G=(V,A)\) be a finite loopless directed graph, with
\(n:=|V|\) and \(m:=|A|\).  A distinguished set \(B\subseteq V\)
represents altruistic donors; we assume throughout that every vertex of
\(B\) is a source.  For a directed path or cycle \(D\), write
\begin{equation}
    \len(D):=|E(D)|.
\end{equation}
Thus, a path of length \(r\) contains \(r+1\) vertices, whereas a directed
cycle of length \(r\) contains \(r\) vertices.

A directed path is \emph{feasible} if it starts at an altruistic donor and
has length at most \(\ell_p\).  A directed cycle is feasible if it contains
no altruistic vertex and has length at most \(\ell_c\).  A
\emph{feasible path--cycle packing} \(\mathcal D\) is a collection of
pairwise vertex-disjoint feasible paths and cycles.

For such a packing, define its transplant cardinality by
\begin{equation}
    c(\mathcal D)
      := \sum_{D\in\mathcal D}\len(D).
\end{equation}
This quantity is exactly the number of recipients receiving a transplant:
each arc of an altruist-initiated chain corresponds to one transplant, and
a directed cycle of length \(r\) performs exactly \(r\) transplants.

An instance of \KEP{} is
\begin{equation}
    I=(G,B,\ell_p,\ell_c,t),
\end{equation}
where \(t\ge1\) is the target transplant cardinality.  The instance is a
yes-instance if there exists a feasible packing \(\mathcal D\) such that
\begin{equation}
    c(\mathcal D)\ge t.
\end{equation}
This is the parameterization and counting convention used by Tian and Xiao
\cite{TianXiao2026}.  Since no recipient can receive more than one
transplant, an instance with \(t>|V\setminus B|\) can be rejected
immediately; hence we may assume \(t\le n\) whenever convenient.

\paragraph{Weighted variant.}
For the optimization extension, each compatibility arc
\(e\in A\) carries an integer utility \(w(e)\), encoded in a machine word.
The additive utility of a feasible packing is
\begin{equation}
    W(\mathcal D)
      := \sum_{e\in E(\mathcal D)} w(e).
\end{equation}
In \emph{Exact-\(t\) Weighted KEP}, the objective is to find, among all
feasible packings satisfying \(c(\mathcal D)=t\), one maximizing
\(W(\mathcal D)\).  Our weighted theorem concerns this exact-cardinality
problem; it does not assert optimality over all packings with
\(c(\mathcal D)\ge t\).

For a family \(\mathcal F\subseteq2^V\) and a set \(S\subseteq V\), define
the disjoint insertion operation
\begin{equation}
    \mathcal F\uplus S
      :=
    \{X\cup S :
       X\in\mathcal F,\ X\cap S=\varnothing\}.
\end{equation}
When \(S=\{x\}\), we also call this an \emph{element insertion}.
We write \(O^*(f(t))\) for \(f(t)\) times a factor polynomial in the input
size.

\subsection{Dynamic Representative Sets}

We use Nederlof's Dynamic Representative Set framework
\cite{Nederlof2026}.  The Boolean interface needed for the decision
algorithm can be stated as follows.

\begin{theorem}[Dynamic Representative Sets {\cite{Nederlof2026}}]
\label{thm:drs-black-box}
Let \(U\) be a universe and let \(0<K<|U|\).  A set family
\(\mathcal F\subseteq2^U\) can be represented dynamically after
preprocessing in
\begin{equation}
    2^{K+O(\sqrt K\log^2 K)}|U|\log|U|
\end{equation}
time.  The representation supports family union and element convolution,
and each such update costs
\begin{equation}
    2^{K+O(\sqrt K\log^2 K)}\log|U|.
\end{equation}
It also supports a query on \(Y\subseteq U\) that returns true exactly when
there exists \(X\in\mathcal F\) satisfying
\begin{equation}
    X\cap Y=\varnothing
    \qquad\text{and}\qquad
    |X|\le K-|Y|.
\end{equation}
Moreover, Weighted Directed \(q\)-Path with word-size edge weights can be
solved deterministically in
\begin{equation}
    2^{q+O(\sqrt q\log^2 q)}(n+m)\log n
\end{equation}
time.
\end{theorem}

The representation need not itself be an explicit subfamily of
\(\mathcal F\); only its disjoint-extension behavior is relevant to our
dynamic program.

Nederlof's framework is formulated more generally over additively
idempotent semirings.  We use this only once, for the weighted extension.
Over the min-plus semiring, a represented table assigns a cost to each
set \(X\), family union takes the minimum of competing costs, and element
convolution propagates the cost while inserting the new element.  A query on
\(Y\) returns
\begin{equation}
    \min\bigl\{
       f(X):
       X\cap Y=\varnothing,\;
       |X|\le K-|Y|
    \bigr\},
\end{equation}
with value \(+\infty\) if no such \(X\) exists.  The asymptotic dependence
on \(K\) is the same as in the Boolean specialization.  Section~\ref{sec:main}
uses precisely this min-plus interface.

\paragraph{Global universe and cardinality budget.}
All DRS states in our Kidney Exchange algorithm use one common universe and
one common cardinality budget.  We pad the graph vertex set with dummy
elements,
\begin{equation}
    U:=V\mathbin{\dot\cup}D,
    \qquad
    |D|:=\max\{0,\,2t+1-|V|\},
\end{equation}
where elements of \(D\) are never inserted into a state family and never
appear in a query.  We then set
    $K:=2t$,
so that \(K<|U|\), as required by Theorem~\ref{thm:drs-black-box}.
Padding therefore has no effect on any disjointness relation involving
actual graph vertices.  The reason the single budget \(K=2t\) suffices for
every relevant Kidney Exchange state is established in
Section~\ref{sec:drs}.

%% file: sections/04_slc.tex
\section{Shortest Long Directed Cycles}
\label{sec:slc}

Before the bounded-component dynamic program can be applied, we must handle
the possibility that one admissible component already contains at least
\(t\) transplants.  Altruist-initiated paths can be truncated, so the
large-path case reduces to an exact-length path test.  Directed cycles are
different: a cycle exceeding the admissible cap cannot be truncated while
remaining a cycle.  Consequently, when \(\ell_c\ge t\), the relevant
quantity is the minimum size of a directed cycle whose size is at least
\(t\).

\begin{definition}[Shortest long-cycle value]
\label{def:lambda}
For a directed graph \(G\) and an integer \(k\ge2\), define
\begin{equation}
    \lambda_k(G)
      :=
    \min\bigl\{
        |V(C)| :
        C\text{ is a simple directed cycle and }|V(C)|\ge k
    \bigr\},
\label{eq:lambda-def}
\end{equation}
with \(\lambda_k(G):=\infty\) if no such cycle exists.
\end{definition}

\begin{theorem}[Shortest Long Directed Cycle]
\label{thm:slc}
Given a directed graph \(G\) and an integer \(k\ge2\),
\(\lambda_k(G)\) can be computed deterministically in
\begin{equation}
    2^{2k+O(\sqrt k\log^2 k)}n^{O(1)}
    =
    O^*(4^{k+o(k)})
\label{eq:slc-runtime}
\end{equation}
time.
\end{theorem}

The algorithm has two regimes.  We first test every cycle size in
\([k,2k]\) exactly using Nederlof's deterministic Weighted Directed Path
algorithm \cite{Nederlof2026}.  If none exists, every long cycle has more
than \(2k\) vertices, and we enter the difficult regime of Zehavi's Long
Directed Cycle framework \cite{Zehavi2016}.  Unlike the decision reduction,
we retain the minimum candidate produced over all compatible partitions.

\begin{algorithm}[t]
\caption{Shortest Long Directed Cycle}
\label{alg:slc}
\begin{algorithmic}[1]
\Require Directed graph \(G=(V,A)\); integer \(k\ge2\)
\Ensure \(\lambda_k(G)\), or \(\infty\) if no such cycle exists
\State \(L^\star\gets\infty\)
\For{\(r=k,\ldots,2k\)}
  \For{each \((u,v)\in A\)}
    \If{there is a simple directed \(v\to u\) path on exactly \(r\) vertices}
      \State \(L^\star\gets\min\{L^\star,r\}\)
    \EndIf
  \EndFor
\EndFor
\If{\(L^\star<\infty\)}
  \State \Return \(L^\star\)
\EndIf
\For{each Zehavi partition \((L,R)\)}
  \For{each ordered pair of distinct vertices \((v,u)\in L\times L\)}
    \State \(P\gets\) a shortest directed \(v\to u\) path in \(G[L]\)
    \If{\(|V(P)|=k\)}
      \State \(H\gets G[V\setminus(V(P)\setminus\{v,u\})]\)
      \State \(P'\gets\) a shortest directed \(u\to v\) path in \(H\)
      \If{\(P'\) exists}
        \State
        \(L^\star\gets
          \min\{L^\star,|V(P)|+|V(P')|-2\}\)
      \EndIf
    \EndIf
  \EndFor
\EndFor
\State \Return \(L^\star\)
\end{algorithmic}
\end{algorithm}

\paragraph{Exact-size regime.}
Fix \(r\in[k,2k]\) and a closing arc \((u,v)\in A\).  It suffices to test
whether \(G\) contains a simple directed \(v\to u\) path on exactly \(r\)
vertices.  We reduce this endpoint-constrained test to Weighted Directed
Path.

Add a new source \(s\) and a new sink \(z\), with new arcs
$
    s\to v$
    and $
    u\to z.
$
Give these two arcs weight zero and every original arc weight one.  Among
simple paths on exactly \(r+2\) vertices in the augmented graph, the minimum
possible weight is \(r-1\), and this value is attained if and only if both
zero-weight arcs are used.  Since \(s\) is a source and \(z\) is a sink,
such a path must have the form
$
    s\to v\rightsquigarrow u\to z,
$
where the middle segment is a simple directed \(v\to u\) path on exactly
\(r\) vertices.  Together with the closing arc \((u,v)\), this segment forms
a directed cycle on exactly \(r\) vertices.  Conversely, every such cycle
gives a path of weight \(r-1\) in the augmented graph.

Nederlof's algorithm therefore gives an exact-\(r\) test with path parameter
\(r+2\le2k+2\).  Testing all \(r\in[k,2k]\) and all closing arcs costs
\begin{equation}
    2^{2k+O(\sqrt k\log^2 k)}n^{O(1)}.
\label{eq:slc-exact-stage}
\end{equation}
If any test succeeds, the smallest successful \(r\) is already
\(\lambda_k(G)\).

\paragraph{The difficult case.}
Assume henceforth that \(G\) contains no directed cycle whose size lies in
\([k,2k]\).  Thus, if \(\lambda_k(G)<\infty\), every shortest long cycle has
more than \(2k\) vertices.

Zehavi constructs, deterministically in
\(4^{k+o(k)}n^{O(1)}\) time, a universal family of bipartitions
\((L,R)\) with the following property: for every ordered sequence of
\(2k\) distinct vertices, some partition places the first \(k\) vertices in
\(L\) and the next \(k\) vertices in \(R\)
\cite{Zehavi2016}.  For a fixed partition and ordered pair
\((v,u)\in L\times L\), the corresponding iteration is polynomial: compute a
shortest directed \(v\to u\) path \(P\) in \(G[L]\), retain it only if
\(|V(P)|=k\), delete its internal vertices, and compute a shortest directed
return path \(P'\) from \(u\) to \(v\).  Whenever \(P'\) exists, the two
paths form a simple directed cycle.

For decision, it suffices that some compatible partition produces a long
cycle.  Recovering the optimum value requires more: one successful
iteration must produce a cycle whose size is exactly \(\lambda_k(G)\).
We therefore fix a globally shortest long cycle first and then choose a
partition compatible with that optimum.

\begin{lemma}[Optimum-compatible partition]
\label{lem:opt-compatible}
Assume that \(G\) contains no directed cycle of size in \([k,2k]\), and let
\begin{equation}
    C^\star=v_1v_2\cdots v_rv_1,
    \qquad
    r=\lambda_k(G)>2k.
\label{eq:opt-cycle}
\end{equation}
Let \((L^\star,R^\star)\) be a Zehavi partition satisfying
\begin{equation}
    \{v_1,\ldots,v_k\}\subseteq L^\star,
    \qquad
    \{v_{k+1},\ldots,v_{2k}\}\subseteq R^\star.
\label{eq:compatible-partition}
\end{equation}
Then the iteration of Algorithm~\ref{alg:slc} associated with
\((v,u)=(v_1,v_k)\) produces a candidate cycle of size exactly \(r\).
\end{lemma}

\begin{proof}[Proof sketch]
Let \(P\) be a shortest directed \(v_1\to v_k\) path in
\(G[L^\star]\).  The prefix
$
    v_1v_2\cdots v_k
$
of \(C^\star\) lies in \(G[L^\star]\), so \(|V(P)|\le k\).  Zehavi's
shortcut argument implies that strict inequality would yield a simple
directed cycle containing at least \(k\) but fewer than \(r\) vertices,
contradicting the minimality of \(C^\star\).  Hence,
\begin{equation}
    |V(P)|=k.
\label{eq:opt-path-size}
\end{equation}
The second shortcut property needed from the same difficult-case argument is
that, after deleting the internal vertices of \(P\), there remains a
directed \(v_k\to v_1\) path \(Q\) satisfying
\begin{equation}
    |V(P)|+|V(Q)|-2\le r.
\label{eq:surviving-return}
\end{equation}
A detailed reconstruction of these shortcut properties is given in the
appendices.
Algorithm~\ref{alg:slc} chooses a shortest surviving return path \(P'\), so
\begin{equation}
    |V(P)|+|V(P')|-2\le r.
\label{eq:candidate-upper}
\end{equation}
Because the internal vertices of \(P\) are deleted before \(P'\) is
computed, \(P\cup P'\) forms a simple directed cycle \(C_P\).  Moreover,
\(|V(P)|=k\), hence \(|V(C_P)|\ge k\).  By the global minimality of
\(C^\star\),
$
    |V(C_P)|\ge\lambda_k(G)=r.
$
Combining this inequality with \eqref{eq:candidate-upper} gives
$
    |V(C_P)|=r.
$
\end{proof}

\begin{proof}[Proof of Theorem~\ref{thm:slc}]
If the exact-size stage succeeds, it returns the smallest cycle size in
\([k,2k]\), which is necessarily \(\lambda_k(G)\).

Suppose instead that all exact-size tests fail.  If no long cycle exists, no
partition-stage iteration produces a candidate, and the algorithm returns
\(\infty\).  Otherwise,  fix a globally shortest long cycle
$
    C^\star=v_1v_2\cdots v_rv_1.
$
Since the exact-size stage has ruled out every size in \([k,2k]\), we have
\(r>2k\).  Zehavi's universal family contains a partition compatible with
the first \(2k\) vertices of \(C^\star\).  By
Lemma~\ref{lem:opt-compatible}, the corresponding iteration produces a
candidate of size exactly
$
    r=\lambda_k(G).
$
Conversely, every candidate produced by Algorithm~\ref{alg:slc} is a simple
directed cycle with at least \(k\) vertices, and therefore has size at least
\(\lambda_k(G)\).  Taking the minimum over all successful iterations returns
exactly \(\lambda_k(G)\).

The exact-size stage costs
$
    2^{2k+O(\sqrt k\log^2 k)}n^{O(1)},
$
while the partition stage costs
$
    4^{k+o(k)}n^{O(1)}.
$
The latter is subsumed by the former, proving
\eqref{eq:slc-runtime}.
\end{proof}

Zehavi's reduction together with Nederlof's path algorithm yields the
corresponding \(O^*(4^{k+o(k)})\) decision bound
\cite{Zehavi2016,Nederlof2026}.  Here we additionally recover the minimum
feasible size \(\lambda_k(G)\) at the same leading exponential base, which
is needed to enforce an upper cycle-length cap in Kidney Exchange.

\begin{lemma}[Large-component reduction]
\label{lem:large-components}
There is a deterministic
\begin{equation}
     2^{2t+O(\sqrt t\log^2 t)}n^{O(1)}
\label{eq:large-component-runtime}
\end{equation}
procedure that either certifies that a \KEP{} instance is a yes-instance or
produces an equivalent instance satisfying
$
     \ell_p<t$
     and 
     $\ell_c<t.
$
\end{lemma}

\begin{proof}
Assume first that \(\ell_p\ge t\).  Add a new source \(s\) and, for every
altruist \(b\in B\), an arc \(s\to b\).  Give the new arcs weight zero and
every original arc weight one.  We ask for a minimum-weight simple directed
path on exactly \(t+2\) vertices.  Such a path has \(t+1\) arcs and weight
at least \(t\).  Equality holds exactly when the path uses one new source
arc followed by \(t\) original arcs.  Since \(s\) has no incoming arcs and
its outgoing neighbors are precisely the altruists, this is equivalent to an
altruist-initiated path containing exactly \(t\) exchange edges.

Any feasible altruist-initiated path longer than \(t\) has a feasible prefix
of exactly \(t\) exchange edges.  Hence, if the test succeeds, the path alone
certifies a yes-instance.  If it fails, no single feasible path can reach
the target, and replacing
$
    \ell_p\leftarrow t-1
$
preserves the answer.  Nederlof's weighted path algorithm performs this test
in
$
    2^{t+O(\sqrt t\log^2 t)}n^{O(1)}
$
time.

Now, assume that \(\ell_c\ge t\).  Unlike a path, a directed cycle cannot in
general be shortened by taking a prefix.  Compute \(\lambda_t(G)\) using
Theorem~\ref{thm:slc}.  Since altruists are sources, no directed cycle
contains a vertex of \(B\).  Therefore, an admissible single cycle covering
at least \(t\) recipients exists exactly when
\begin{equation}
    \lambda_t(G)\le\ell_c.
\label{eq:large-cycle-test}
\end{equation}
If so, that cycle alone certifies a yes-instance.  Otherwise, no admissible
cycle has size in \([t,\ell_c]\), so replacing
$
    \ell_c\leftarrow t-1
$
preserves the answer.

After the two tests, either a yes-instance has been certified, or both caps
are strictly smaller than \(t\).  The path branch is below the claimed
running time, while the cycle branch is bounded by
Theorem~\ref{thm:slc}.  Hence, the whole reduction runs in
\eqref{eq:large-component-runtime}.
\end{proof}

%% file: sections/05_semifeasible.tex
\section{The Semi-Feasible Dynamic Program}
\label{sec:semifeasible}

By Lemma~\ref{lem:large-components}, we may henceforth assume
$
    \ell_p<t$
    and 
    $\ell_c<t.$
In this bounded-component case, we use the semi-feasible dynamic-programming
formulation of Tian and Xiao \cite{TianXiao2026}.  The purpose of this
section is to expose the semantic state space and, in particular, the
restricted algebra by which its used-vertex families evolve.  Section
\ref{sec:drs} will show that this algebra can be executed directly by
Dynamic Representative Sets.

\subsection{Semantic states}

A semi-feasible packing consists of completed feasible paths and cycles
together with one distinguished active component.  Let
\(D_\ell^{vu}\) denote an active component of exchange length \(\ell\)
with distinguished vertices \(v\) and \(u\).  For integers
\(k,\ell,p\) and vertices \(v,u\), let
$
    \mathcal F_{k,\ell,p}^{vu}
$
denote the family of used-vertex sets arising from all semi-feasible
packings having total exchange length \(k\), active component
\(D_\ell^{vu}\), and exactly \(p\) used vertices.  Thus, 
\begin{equation}
    X\in\mathcal F_{k,\ell,p}^{vu}
    \quad\Longrightarrow\quad
    |X|=p.
\end{equation}
The indices \((k,\ell,p,v,u)\) retain the state-local information needed to
decide which transitions are legal.  Once these indices are fixed, the
family \(\mathcal F_{k,\ell,p}^{vu}\) needs to record only which vertices
have already been used.  This separation between local state information
and global vertex-disjointness is the feature that will permit a dynamic
set-family representation.

We write \(\mathsf{Term}(v,u,\ell)\) when the active component is a feasible
completed component.  Concretely, this holds when either
$
    v\in B$,  $u\notin B$, and $\ell\le\ell_p$,
so that the active component is an altruist-initiated path, or when
$
    v=u\notin B$ and $\ell\le\ell_c$,
so that it is a directed cycle.

\subsection{Recurrence skeleton and set-family algebra}

We now record the part of Tian and Xiao's recurrence that is relevant to the
representation of used vertices.  The full transition guards are those of
their semantic dynamic program \cite{TianXiao2026}; the appendices
material records them explicitly.  After those guards are fixed, every
candidate family is obtained by one of the following four forms.

A new active edge initializes a state with the two incident vertices,
\begin{equation}
    \mathcal N_{k,\ell,p}^{vu}
       = \{\{v,u\}\}.
    \label{eq:tx-init}
\end{equation}
A completed component may be followed by a new active component, which adds
two previously unused vertices:
\begin{align}
\mathcal N_{k,\ell,p}^{vu}
={}&
 \bigcup_{\substack{v'\in B,\ u'\notin B\\
                    1\le\ell'\le\ell_p}}
 \Bigl(
   \mathcal F_{k-1,\ell',p-2}^{v'u'}
   \uplus\{v,u\}
 \Bigr)
 \nonumber\\[-2mm]
&\quad\cup
 \bigcup_{\substack{v'\notin B\\
                    1\le\ell'\le\ell_c}}
 \Bigl(
   \mathcal F_{k-1,\ell',p-2}^{v'v'}
   \uplus\{v,u\}
 \Bigr).
\label{eq:tx-start}
\end{align}
Extending the active component by one exchange edge inserts one new
vertex:
\begin{equation}
\mathcal N_{k,\ell,p}^{vu}
 =
 \bigcup_{\substack{w\ne v\\(w,u)\in A}}
 \Bigl(
   \mathcal F_{k-1,\ell-1,p-1}^{vw}
   \uplus\{u\}
 \Bigr).
\label{eq:tx-extend}
\end{equation}
Finally, closing an active path into a cycle adds an exchange edge but no
new used vertex:
\begin{equation}
\mathcal N_{k,\ell,p}^{vv}
 =
 \bigcup_{\substack{w\ne v\\(w,v)\in A}}
 \mathcal F_{k-1,\ell-1,p}^{vw}.
\label{eq:tx-close}
\end{equation}
Whenever the corresponding semantic guard fails, the contribution of that
case is empty.

The relevant point is the set-family structure: initialization inserts two vertices; starting a component applies two insertions and union; extension applies one insertion and union; and cycle closing uses union only.  The corresponding cardinality changes are $0\to2$, $p-2\to p$, $p-1\to p$, and $p\to p$, respectively.

In particular, the recurrence never combines two independently evolving
used sets, deletes a used vertex, or performs arbitrary subset convolution.
This restricted update algebra is precisely the interface supported by the
dynamic representation introduced in Section~\ref{sec:drs}.

\subsection{Residual extension budget}

The cardinality index \(p\) plays a structural role.  Future components must
avoid all vertices already used by the current partial packing, so a state
must preserve enough information to answer bounded disjoint-extension
queries.  Tian and Xiao's representative-family construction preserves, for
a state containing \(p\) used vertices, all disjointness queries against
future vertex sets \(Y\) satisfying
\begin{equation}
    |Y|\le 2t-p.
\label{eq:tx-residual-budget}
\end{equation}
Equivalently, their state family is maintained with a representative-set
parameter \(2t-p\) \cite{TianXiao2026}.

This state-dependent quantity is exactly what allows all states to be placed
under a single global DRS budget.  Setting
\begin{equation}
    K:=2t,
\label{eq:global-drs-budget}
\end{equation}
a state of cardinality \(p\) has residual disjoint-extension budget
\begin{equation}
    K-p=2t-p.
\label{eq:state-residual-budget}
\end{equation}
Moreover, the recurrence respects this budget locally.  Suppose a transition
inserts a fixed set \(S\) of \(s\) vertices and takes a predecessor of
cardinality \(p-s\) to a successor of cardinality \(p\).  If a future set
\(Y\) satisfies
$
      |Y|\le K-p,
$
then, whenever \(S\cap Y=\varnothing\),
$
      |Y\cup S|
      \le K-p+s
      = K-(p-s).
$
Thus, the successor's admissible extension query becomes an admissible query
for the predecessor after adjoining precisely the vertices inserted by the
transition.  Section~\ref{sec:drs} turns this residual-budget identity into
the formal statewise extension invariant.

Finally, we use the bounded-witness structure of Tian and Xiao.  In the
bounded-component regime, every yes-instance has a relevant accepting
semi-feasible state satisfying $
    t\le k\le p\le2t
$
\cite{TianXiao2026}.  Tracing a constructive recurrence derivation backward
from such an accepting state preserves the same bound: the recurrence only
inserts used vertices and never deletes them, so every predecessor used set
is a subset of the terminal used set.  Hence every state on the accepting
derivation has cardinality at most \(2t\).  Consequently, the entire
derivation can be maintained under the single global DRS budget \(K=2t\).

%% file: sections/06_drs.tex
\section{Dynamic Representative Implementation}
\label{sec:drs}

The recurrence of Section~\ref{sec:semifeasible} describes exact semantic
families of used vertices.  We now show that these families need never be
materialized.  Instead, the entire recurrence can be executed inside
Nederlof's Dynamic Representative Set framework under the single global
cardinality budget
$
    K=2t.
$
All graph-theoretic information (edge incidence, endpoints, altruist
status, component length, and transition legality) remains in the semantic
state indices.  The dynamic representation stores only the information
needed to preserve future vertex-disjoint extensions.

\subsection{Statewise extension equivalence}

For a semantic state
\begin{equation}
    \sigma=(k,\ell,p,v,u),
\end{equation}
write
\begin{equation}
    \mathcal F_\sigma
      :=\mathcal F_{k,\ell,p}^{vu}.
\end{equation}
Every member of \(\mathcal F_\sigma\) has cardinality exactly \(p\).

\begin{definition}[Statewise extension equivalence]
\label{def:extension-equivalence}
Let \(\sigma=(k,\ell,p,v,u)\) with \(p\le2t\).
A maintained DRS object \(\mathfrak R_\sigma\) is
\emph{extension-equivalent} to \(\mathcal F_\sigma\) if, for every
\(Y\subseteq V\) satisfying
\begin{equation}
      |Y|\le2t-p,
\end{equation}
we have
\begin{equation}
  \Query(\mathfrak R_\sigma,Y)=1
  \quad\Longleftrightarrow\quad
  \exists X\in\mathcal F_\sigma
  \text{ such that }X\cap Y=\varnothing.
\end{equation}
\end{definition}
The definition matches the formal DRS query interface exactly.  Indeed,
under the global budget \(K=2t\), every \(X\in\mathcal F_\sigma\) has
\(|X|=p\).  For every admissible query \(Y\), we have
\begin{equation}
      |Y|\le K-p,
\end{equation}
and therefore
\begin{equation}
      |X|=p\le K-|Y|.
\end{equation}
Thus, the cardinality condition in
Theorem~\ref{thm:drs-black-box} is automatic, while the remaining query
condition is precisely disjointness from \(Y\).  The state-dependent
residual extension budget is therefore
\begin{equation}
      K-p=2t-p.
\label{eq:state-residual-extension-budget}
\end{equation}

\subsection{Residual-budget transfer}

The compatibility between the recurrence and DRS can be stated
independently of Kidney Exchange.

\begin{lemma}[Residual-budget transfer]
\label{lem:transfer}
Fix a global cardinality budget \(K\).  Let a successor family of \(p\)-sets
be a union of terms of the form
\begin{equation}
      \mathcal F'\uplus S,
      \qquad |S|=s,
\end{equation}
where every member of \(\mathcal F'\) has cardinality \(p-s\), and \(S\) is
a fixed set of inserted elements.  Suppose a maintained representation of
\(\mathcal F'\) is correct for every disjoint-extension query of size at
most
\begin{equation}
      K-(p-s).
\end{equation}
Then applying the \(s\) corresponding element convolutions and taking DRS
unions preserves every successor disjoint-extension query of size at most
\begin{equation}
      K-p.
\end{equation}
\end{lemma}

\begin{proof}
Fix a query set \(Y\) with \(|Y|\le K-p\).

If \(S\cap Y\ne\varnothing\), then every member of
\(\mathcal F'\uplus S\) intersects \(Y\), so this transition contributes no
feasible successor to the query.

Assume now that \(S\cap Y=\varnothing\).  For every
\(X\in\mathcal F'\),
\begin{equation}
      (X\cup S)\cap Y=\varnothing
      \quad\Longleftrightarrow\quad
      X\cap(Y\cup S)=\varnothing.
\end{equation}
Moreover,
\begin{equation}
      |Y\cup S|
      \le |Y|+|S|
      \le K-p+s
      = K-(p-s).
\end{equation}
Hence \(Y\cup S\) is a valid predecessor query.  Correctness of the
predecessor representation therefore gives exactly the correct answer for
this transition term.  Finally, existence over a union of semantic families
is Boolean disjunction, which is exactly the effect of DRS family union.
\end{proof}

The identity
\begin{equation}
      K-(p-s)=(K-p)+s
\end{equation}
is the key point.  Inserting \(s\) vertices consumes exactly \(s\) units of
used-set cardinality and simultaneously enlarges the predecessor's available
extension budget by exactly \(s\).  Thus the disjointness obligation created
by the inserted set is absorbed without increasing the global parameter
\(K\).

For the Kidney Exchange recurrence, \(K=2t\), and the three non-base
cardinality changes in
Equations~\eqref{eq:tx-start}--\eqref{eq:tx-close} correspond exactly to
\begin{equation}
      s=2,\qquad s=1,\qquad s=0.
\end{equation}

\begin{corollary}[Closure of the Kidney Exchange recurrence]
\label{cor:recurrence-compatible}
Assume that all predecessor states of exchange length \(k-1\) and
cardinality at most \(2t\) are represented extension-equivalently.
Then every valid successor state of exchange length \(k\) and cardinality
at most \(2t\) obtained from
Equations~\eqref{eq:tx-start}--\eqref{eq:tx-close}
can also be represented extension-equivalently using only DRS union and
element convolution.
\end{corollary}

\begin{proof}
Apply Lemma~\ref{lem:transfer} to each legal recurrence term:
\(s=2\) when starting a new component, \(s=1\) when extending the active
component, and \(s=0\) when closing a cycle.  Alternative predecessor terms
are combined by DRS union.  Initialization is obtained from the singleton
family \(\{\varnothing\}\) by the required element insertions.
All graph-theoretic legality conditions are enforced by the semantic state
indices and therefore require no additional set-family operation.
\end{proof}

\subsection{Global statewise invariant}

The previous result is local to one transition.  We now lift it to the whole
dynamic program.
\begin{proposition}[Statewise extension invariant]
\label{prop:statewise-invariant}
For every semantic state
\begin{equation}
      \sigma=(k,\ell,p,v,u)
      \qquad\text{with }p\le2t
\end{equation}
that is constructed by the bounded-component recurrence, the algorithm
maintains a DRS object \(\mathfrak R_\sigma\) extension-equivalent to the
exact semantic family \(\mathcal F_\sigma\).
\end{proposition}
\begin{proof}
We induct on the exchange length \(k\).
For an initialization state, the exact semantic family is obtained from
\(\{\varnothing\}\) by inserting the vertices prescribed by the base
transition.  The DRS representation performs the same supported element
convolutions, and therefore represents the base family correctly.

Assume the claim holds for every relevant state of exchange length \(k-1\).
Every state at layer \(k\) is the union of legal transition terms described
in Section~\ref{sec:semifeasible}.  By the induction hypothesis, each
predecessor is extension-equivalent to its semantic family.
Corollary~\ref{cor:recurrence-compatible} then preserves extension
equivalence through every transition term and through their union.
Therefore, the claim holds for every state at layer \(k\).
\end{proof}
This proposition is the point at which the static and dynamic approaches
separate algorithmically.  Both preserve the same future-extension
semantics.  A static implementation first constructs a candidate family and
then compresses it to a representative subfamily.  Here, by contrast, the
candidate-family recurrence itself is evaluated inside a representation
already closed under its update algebra.  No state requires explicit
enumeration of the candidate family before compression.

\subsection{The global \(2t\)-vertex trajectory}

It remains to justify that a single DRS budget \(K=2t\) covers every state
needed by an accepting computation.
\begin{lemma}[Bounded accepting trajectory]
\label{lem:bounded-trajectory}
If the bounded-component instance is a yes-instance, then the refined
semantic recurrence has an accepting derivation in which every state has
used-vertex cardinality at most \(2t\).
\end{lemma}

\begin{proof}
Tian and Xiao's bounded-cardinality result yields an accepting semantic
family member
\begin{equation}
      X^\star,
      \qquad
      |X^\star|=p^\star\le2t
\end{equation}
\cite{TianXiao2026}.  Fix one actual semi-feasible packing represented by
\(X^\star\), and trace a constructive recurrence derivation of this packing
backwards.

Along every forward transition, the recurrence inserts zero, one, or two
previously unused vertices and never removes a used vertex.  Hence every
predecessor used-set on the chosen derivation is a subset of \(X^\star\).
Consequently, every state on the derivation has cardinality at most
$
      |X^\star|\le2t.
$
\end{proof}
Combining Proposition~\ref{prop:statewise-invariant} with
Lemma~\ref{lem:bounded-trajectory} gives the required completeness
statement.  At an accepting terminal state, querying with
$
      Y=\varnothing
$
yields
\begin{equation}
  \Query(\mathfrak R_\sigma,\varnothing)=1
  \quad\Longleftrightarrow\quad
  \mathcal F_\sigma\ne\varnothing,
\end{equation}
because \(p\le K=2t\).  Thus, the dynamic representation loses no accepting
witness.

\subsection{The bounded-component algorithm}

Algorithm~\ref{alg:drs-kep} summarizes the resulting computation.  The
semantic layer determines which transitions are legal, while the DRS layer
executes only the corresponding set-family operations.

\begin{algorithm}[t]
\caption{DRS-KEP in the bounded-component regime}
\label{alg:drs-kep}
\begin{algorithmic}[1]
\Require A KEP instance with \(\ell_p<t\) and \(\ell_c<t\)
\Ensure YES iff a feasible packing covers at least \(t\) recipients
\State initialize the padded universe and set \(K\gets2t\)
\For{\(k=1,\ldots,2t\)}
  \For{each valid semantic state \((k,\ell,p,v,u)\) with \(p\le2t\)}
    \If{\((k,\ell,p,v,u)\) is a legal base state}
      \State initialize \(\mathfrak R_{k,\ell,p}^{vu}\)
    \Else
      \State construct \(\mathfrak R_{k,\ell,p}^{vu}\) from its legal
             predecessor terms
      \Statex \hspace{1.2em} using DRS union and at most two element
             convolutions per term
    \EndIf
  \EndFor
\EndFor
\For{each valid state \((k,\ell,p,v,u)\) with
      \(t\le k\le2t\) and \(p\le2t\)}
  \If{\(\mathsf{Term}(v,u,\ell)\) and
      \(\Query(\mathfrak R_{k,\ell,p}^{vu},\varnothing)=1\)}
    \State \Return YES
  \EndIf
\EndFor
\State \Return NO
\end{algorithmic}
\end{algorithm}

The predicate \(\mathsf{Term}(v,u,\ell)\) is exactly the terminal-component
condition defined in Section~\ref{sec:semifeasible}.  Hence acceptance
introduces no additional DRS semantics: querying with the empty future set
tests precisely whether the corresponding exact semantic family is
nonempty.

\begin{lemma}[Running time]
\label{lem:dp-runtime}
Algorithm~\ref{alg:drs-kep} runs deterministically in
\begin{equation}
      2^{2t+O(\sqrt t\log^2 t)}n^{O(1)}
      =
      O^*(4^{t+o(t)}).
\label{eq:bounded-dp-runtime}
\end{equation}
\end{lemma}

\begin{proof}
The state indices satisfy
\begin{equation}
      1\le k,p\le2t,
      \qquad
      1\le\ell<t,
      \qquad
      (v,u)\in V^2.
\label{eq:bounded-state-range}
\end{equation}
Thus, there are \(t^{O(1)}n^2\) semantic states.  For each fixed state, the
recurrence enumerates only polynomially many predecessor endpoint choices
and active lengths.  Each recurrence term uses at most two element
convolutions, and the alternative terms are accumulated by a polynomial
number of DRS unions.  Hence the entire dynamic program performs
$
      t^{O(1)}n^{O(1)}
$
DRS operations.
With the global budget \(K=2t\),
Theorem~\ref{thm:drs-black-box} gives per-operation time
\begin{equation}
      2^{2t+O(\sqrt t\log^2 t)}\log|U|,
\label{eq:drs-operation-cost}
\end{equation}
while preprocessing takes
\begin{equation}
      2^{2t+O(\sqrt t\log^2 t)}
      |U|\log|U|.
\label{eq:drs-preprocessing-cost}
\end{equation}
The padded universe satisfies
\begin{equation}
      |U|=O(n+t).
\label{eq:padded-universe-size}
\end{equation}
After immediately rejecting instances with \(t>|V\setminus B|\), we have
\(t\le n\).  Therefore, \(|U|=O(n)\), and the universe size, the number of
states, the number of transitions, and logarithmic DRS factors contribute only
polynomial factors.  Combining these bounds yields
\eqref{eq:bounded-dp-runtime}.
\end{proof}

%% file: sections/07_main_theorem.tex
\section{Main Result}
\label{sec:main}

We now assemble the two algorithmic parts.  It is useful to separate the
correctness argument into three interfaces: large-component normalization,
semantic preservation of the bounded DP, and terminal detection.  This also
makes the running-time accounting independent of the details of the
Tian--Xiao state guards.

\begin{theorem}[Deterministic Kidney Exchange]
\label{thm:main}
Kidney Exchange, parameterized by the target number \(t\) of covered
recipients, can be solved deterministically in
\begin{equation}
  2^{2t+O(\sqrt t\log^2t)}n^{O(1)}
\end{equation}
time.  Consequently,
$
   \KEP\in O^*(4^{t+o(t)}).
$
\end{theorem}

\begin{proof}
If \(t>|V\setminus B|\), return no.  Otherwise apply
Lemma~\ref{lem:large-components}.  If the reduction finds a feasible
altruist-initiated path or directed cycle covering at least \(t\) recipients,
return yes.  Otherwise, it produces an equivalent instance satisfying
$
   \ell_p<t$ and $
   \ell_c<t.
$
Run the refined semi-feasible DP of Section~\ref{sec:semifeasible}, maintaining
for every semantic state with \(p\le2t\) the DRS object of
Proposition~\ref{prop:statewise-invariant}.  The semantic recurrence and all
of its legality tests are unchanged; only the way each used-vertex family is
maintained is changed.  By Proposition~\ref{prop:statewise-invariant}, every
maintained object is extension-equivalent to its corresponding semantic
family under the global budget \(K=2t\).

Suppose first that the instance is positive.
Lemma~\ref{lem:bounded-trajectory} gives an accepting derivation lying
entirely within states of cardinality at most \(2t\).  Statewise extension
equivalence is therefore valid throughout that derivation.  At an accepting
terminal state \(\sigma\), Definition~\ref{def:extension-equivalence} with
\(Y=\varnothing\) gives
\begin{equation}
   \Query(\mathfrak R_\sigma,\varnothing)=1
   \quad\Longleftrightarrow\quad
   \mathcal F_\sigma\ne\varnothing.
\end{equation}
Hence, the algorithm accepts.

Conversely, if the algorithm accepts, then some valid terminal state
\(\sigma\) satisfies
\(\Query(\mathfrak R_\sigma,\varnothing)=1\).
By extension equivalence,
\(\mathcal F_\sigma\ne\varnothing\).  The Tian--Xiao terminal-state
characterization therefore yields a genuine feasible packing covering at
least \(t\) recipients.  Thus no false positive is introduced by the DRS
representation, and the algorithm is correct.

The large altruistic-path branch costs
$
   2^{t+O(\sqrt t\log^2t)}n^{O(1)}.
$
By Theorem~\ref{thm:slc}, the large-cycle branch costs
$
   2^{2t+O(\sqrt t\log^2t)}n^{O(1)},
$
and Lemma~\ref{lem:dp-runtime} gives the same bound for the bounded-component
DP.  Hence, the full deterministic running time is
$
  2^{2t+O(\sqrt t\log^2t)}n^{O(1)}.
$
Finally, \(\sqrt t\log^2t=o(t)\), so
$
  2^{2t+O(\sqrt t\log^2t)}
  =
  4^{t+o(t)}.
$
\end{proof}

\paragraph{Where the improvement comes from.}
Tian and Xiao already supply the bounded-cardinality semantic DP; the
improvement is not a smaller witness bound.  Instead, the DRS representation
turns the state budget \(p\le2t\) directly into a \(2^{2t+o(t)}\) dynamic
maintenance cost, avoiding repeated static representative-family
computation.  The SLC theorem ensures that the normalization needed to reach
the bounded-component regime remains within the same leading exponential
base.

\subsection{Cardinality-first weighted optimization}
\label{sec:weighted}

The dynamic representation also preserves an optimization layer useful in
kidney-exchange clearing.  Let
\(w:A\to\mathbb Z_{\ge0}\) be word-size additive arc utilities and define
$
   W(\mathcal D)
      :=
   \sum_{e\in E(\mathcal D)}w(e).
$
In \emph{Exact-\(t\) Weighted Kidney Exchange}, the objective is to maximize
\(W(\mathcal D)\) over feasible packings containing exactly \(t\) transplant
arcs.  The exact-cardinality restriction is important: the bounded-witness
reduction for the decision problem need not preserve maximum utility among
all packings covering at least \(t\) recipients.

\begin{theorem}[Exact-cardinality weighted KEP]
\label{thm:weighted-exact}
Exact-\(t\) Weighted Kidney Exchange can be solved deterministically in
\begin{equation}
  2^{2t+O(\sqrt t\log^2t)}n^{O(1)}
  =
  O^*(4^{t+o(t)}).
\end{equation}
\end{theorem}

\begin{proof}[Proof sketch]
Let
\begin{equation}
   M:=\max\bigl(\{0\}\cup\{w(e):e\in A\}\bigr),
   \qquad
   d(e):=M-w(e).
\end{equation}
For every feasible packing \(\mathcal D\) containing exactly \(t\)
transplant arcs,
\begin{equation}
   \sum_{e\in E(\mathcal D)}d(e)
      =
   tM-W(\mathcal D).
\end{equation}
Thus, minimizing transformed cost is equivalent to maximizing utility.

Nederlof's DRS framework works over additively idempotent semirings,
including the min-plus specialization used for Weighted Directed
\(k\)-Path \cite{Nederlof2026}.  We therefore maintain the semantic
minimum-cost functions through min-plus DRS objects.  Union becomes minimum,
element insertion is handled by element convolution, and each semantic
transition adds the transformed cost of its unique newly introduced
transplant arc.  The residual-budget argument of
Lemma~\ref{lem:transfer} is purely set-theoretic and is unchanged, so the
global budget remains \(K=2t\).

It remains to separate components of length \(t\) from the bounded-component
case.  If an exact-\(t\) solution contains a component of length \(t\), that
component accounts for all transplant arcs and is the only nonempty
component.  For an altruist-initiated chain, add a new source with zero-cost
arcs to the altruists and solve the corresponding source-constrained
exact-\(t\) Weighted Directed Path instance.  For a cycle, enumerate its
closing arc \((u,v)\), optimize an exact-size directed \(v\to u\) path, and
add the cost \(d(u,v)\).  Both cases take
$
   2^{t+O(\sqrt t\log^2t)}n^{O(1)}
$
time.

Otherwise every component has length at most \(t-1\).  An exact-\(t\)
packing uses at most \(2t\) vertices: cycles contribute one used vertex per
transplant arc, while every nonempty altruist-initiated chain contributes
one additional altruist, and there are at most \(t\) such chains.
Consequently, the min-plus semi-feasible recurrence is maintained under the
same global budget \(K=2t\), giving
$
   2^{2t+O(\sqrt t\log^2t)}n^{O(1)}
$
time.  Taking the minimum transformed cost over the exact-\(t\) chain case,
the exact-\(t\) cycle case, and the bounded-component case proves the
theorem.  Appendix~D gives the weighted recurrence and
statewise invariant in detail.
\end{proof}

\begin{corollary}[Cardinality-first additive utility]
\label{cor:weighted-lex}
Let \(\tau\) be the maximum number of recipients that can be covered by a
feasible packing.  A packing lexicographically maximizing
\((c(\mathcal D),W(\mathcal D))\) can be computed deterministically in
$
   O^*(4^{\tau+o(\tau)})
$
time.
\end{corollary}

\begin{proof}
Apply Theorem~\ref{thm:main} successively to targets
\(1,2,\ldots,\tau+1\); the first negative target is \(\tau+1\).
The total decision cost satisfies
\begin{equation}
   \sum_{j=1}^{\tau+1}
      2^{2j+O(\sqrt j\log^2j)}
   =
   O^*(4^{\tau+o(\tau)}),
\end{equation}
since the final call changes the leading exponential term only by a constant
factor.  Having determined \(\tau\), apply
Theorem~\ref{thm:weighted-exact} with \(t=\tau\).
\end{proof}

%% file: sections/08_conclusion.tex
\section{Conclusion}
\label{sec:conclusion}

We obtained a deterministic \(O^*(4^{t+o(t)})\) algorithm for Kidney
Exchange, matching the leading base of the known randomized \(O^*(4^t)\)
bound.  The two ingredients are a deterministic \(4^{k+o(k)}\)
shortest-long-cycle algorithm for large-component normalization and a DRS
implementation of the bounded semi-feasible recurrence under the single
global budget \(K=2t\), enabled by the residual-budget identity.  The same
representation yields exact-cardinality and cardinality-first
additive-utility optimization at the same base.  Natural directions are a
deterministic base below \(4\), matching conditional lower bounds, and
identifying richer clearing objectives compatible with similarly compact
dynamic representations.

%% file: appendices/A_slc_reconstruction.tex
\section{Detailed Reconstruction of the Shortest-Long-Cycle Argument}
\label{app:slc-reconstruction}

This appendix expands the two shortcut arguments used in the difficult
regime of the shortest-long-cycle algorithm.  The underlying
universal-family construction and the two shortcut observations reconstructed
below are due to Zehavi \cite{Zehavi2016}; our additional use is to fix a
globally shortest long cycle and retain the minimum candidate over all
partitions.

\subsection{Exact-size regime}

Fix \(r\in\{k,\ldots,2k\}\) and a closing arc \((u,v)\in A\).  Add fresh
vertices \(s,z\), add only the gadget arcs \(s\to v\) and \(u\to z\), assign
weight zero to these two arcs, and assign weight one to every original arc.
A simple directed path on exactly \(r+2\) vertices has \(r+1\) arcs and
weight at least \(r-1\).  Equality holds exactly when both zero-weight arcs
are used, in which case the path has the form
\[
    s\to v\rightsquigarrow u\to z
\]
and its middle segment is a simple directed \(v\to u\) path on exactly
\(r\) original vertices.  Together with \((u,v)\), this segment forms an
\(r\)-vertex directed cycle, and conversely every such cycle yields a gadget
path of weight \(r-1\).

Nederlof's deterministic Weighted Directed Path algorithm therefore tests a
fixed \(r\) and closing arc in
\[
    2^{r+O(\sqrt r\log^2 r)}n^{O(1)}
\]
time \cite{Nederlof2026}.  Since \(r\le2k\), enumerating all
\(r\in[k,2k]\) and all closing arcs costs
\begin{equation}
    2^{2k+O(\sqrt k\log^2 k)}n^{O(1)}.
    \label{eq:app-slc-exact-runtime}
\end{equation}

\subsection{Difficult regime: reconstruction of Zehavi's shortcuts}

Assume that \(G\) has no directed cycle with size in \([k,2k]\), and fix a
globally shortest long cycle
\begin{equation}
    C^\star=v_1v_2\cdots v_rv_1,
    \qquad
    r=\lambda_k(G)>2k.
    \label{eq:app-opt-cycle}
\end{equation}
Choose a partition \((L^\star,R^\star)\) from Zehavi's deterministic
universal family such that
\begin{equation}
    v_1,\ldots,v_k\in L^\star,
    \qquad
    v_{k+1},\ldots,v_{2k}\in R^\star.
    \label{eq:app-compatible-partition}
\end{equation}
Such a partition exists by the universal-family construction in
\cite{Zehavi2016}.  Let
\[
    P=u_1u_2\cdots u_q
\]
be a shortest directed \(v_1\to v_k\) path in \(G[L^\star]\), where
\(u_1=v_1\) and \(u_q=v_k\).  The prefix
\(v_1v_2\cdots v_k\) of \(C^\star\) lies in \(G[L^\star]\), so
\(q\le k\).

\paragraph{First shortcut: the forward path has exactly \(k\) vertices.}
Suppose for contradiction that \(q<k\).  Traverse the return portion
\[
    v_{k+1},v_{k+2},\ldots,v_r,v_1
\]
of \(C^\star\), and let \(v_i\) be its first vertex belonging to \(V(P)\).
Such a vertex exists because \(v_1\in V(P)\).

If \(v_i=v_1\), then \(P\) together with the entire return portion of
\(C^\star\) forms a simple directed cycle.  Its size is
\[
    q+(r-k+2)-2
    =
    r-(k-q)
    <
    r.
\]
It contains the \(k\) distinct vertices
\(v_{k+1},\ldots,v_{2k}\), and hence is a long cycle, contradicting the
minimality of \(C^\star\).

Otherwise, write \(v_i=u_j\).  Since \(V(P)\subseteq L^\star\) while
\(v_{k+1},\ldots,v_{2k}\in R^\star\), we have \(i>2k\).  By the choice of
\(v_i\), the directed walk
\[
    v_{k+1}\to\cdots\to v_i(=u_j)
    \to u_{j+1}\to\cdots\to u_q(=v_k)\to v_{k+1}
\]
is a simple directed cycle.  It contains
\(v_{k+1},\ldots,v_{2k}\), so it has at least \(k\) vertices.  Moreover,
it replaces the portion
\[
    v_1\to v_2\to\cdots\to v_i
\]
of \(C^\star\) by a subpath of \(P\), and since \(q<k<i\), the resulting
cycle is strictly shorter than \(C^\star\).  This again contradicts the
definition of \(r=\lambda_k(G)\).

Therefore,
\begin{equation}
    |V(P)|=k.
    \label{eq:app-forward-size}
\end{equation}
This is Zehavi's first shortcut observation specialized to the optimum cycle
fixed in \eqref{eq:app-opt-cycle} \cite{Zehavi2016}.

\paragraph{Second shortcut: the return suffix survives.}
Set
\begin{equation}
    H:=
    G\bigl[V\setminus(V(P)\setminus\{v_1,v_k\})\bigr].
    \label{eq:app-residual-graph}
\end{equation}
Suppose for contradiction that some vertex of \(P\) occurs among
\(v_{k+1},\ldots,v_r\), and let \(v_i=u_j\) be the first such vertex along
this suffix.  Since
\[
    V(P)\subseteq L^\star
    \qquad\text{and}\qquad
    v_{k+1},\ldots,v_{2k}\in R^\star,
\]
we have \(i>2k\).  By the choice of \(v_i\),
\[
    v_{k+1}\to\cdots\to v_i(=u_j)
    \to u_{j+1}\to\cdots\to u_k(=v_k)\to v_{k+1}
\]
is a simple directed cycle.  It contains the \(k\) distinct vertices
\(v_{k+1},\ldots,v_{2k}\), and therefore has at least \(k\) vertices.
It is strictly shorter than \(C^\star\), contradicting
\(r=\lambda_k(G)\).

Consequently,
\begin{equation}
    V(P)\cap\{v_{k+1},\ldots,v_r\}=\varnothing.
    \label{eq:app-no-suffix-intersection}
\end{equation}
Hence, the original return suffix
\begin{equation}
    Q:=v_kv_{k+1}\cdots v_rv_1
    \label{eq:app-surviving-suffix}
\end{equation}
is a directed \(v_k\to v_1\) path in \(H\).  Since
\[
    |V(P)|=k
    \qquad\text{and}\qquad
    |V(Q)|=r-k+2,
\]
we obtain
\begin{equation}
    |V(P)|+|V(Q)|-2
    =
    k+(r-k+2)-2
    =
    r.
    \label{eq:app-return-equality}
\end{equation}
Let \(P'\) be a shortest directed \(v_k\to v_1\) path in \(H\).  Since
\(Q\) is available,
\begin{equation}
    |V(P)|+|V(P')|-2\le r.
    \label{eq:app-candidate-upper}
\end{equation}
The internal vertices of \(P\) are absent from \(H\), so \(P\cup P'\)
forms a simple directed cycle \(C_P\).  By
\eqref{eq:app-forward-size}, \(|V(C_P)|\ge k\); hence the global minimality
of \(C^\star\) gives
\[
    |V(C_P)|\ge r.
\]
Together with \eqref{eq:app-candidate-upper}, this yields
\begin{equation}
    |V(C_P)|=r=\lambda_k(G).
    \label{eq:app-opt-candidate}
\end{equation}

\subsection{Recovering the optimum over all partitions}

The universal-family guarantee is existential in the required order: first
fix the globally shortest long cycle \(C^\star\), and then choose a
partition compatible with its first \(2k\) vertices.  The corresponding
iteration produces the candidate in \eqref{eq:app-opt-candidate}.
In contrast, every candidate produced by a successful iteration is a simple
directed cycle with at least \(k\) vertices and therefore has size at least
\(\lambda_k(G)\).  Taking the minimum over all successful iterations thus
returns exactly \(\lambda_k(G)\).

%% file: appendices/B_semifeasible_reconstruction.tex
\section{Semi-Feasible Recurrence and DRS Interface}
\label{app:tx-reconstruction}

For completeness, we record the semi-feasible recurrence of Tian and Xiao
\cite{TianXiao2026} in the notation used in the main paper.  The purpose is
to expose the set-family operations and their cardinality changes; all
path/cycle type and length guards remain the semantic legality conditions of
the original recurrence.

Let \(\mathcal F_{k,\ell,p}^{vu}\) denote the family of used-vertex sets of
semi-feasible packings with total exchange length \(k\), active component
\(D_\ell^{vu}\), and exactly \(p\) used vertices.  Thus
\begin{equation}
    X\in\mathcal F_{k,\ell,p}^{vu}
    \quad\Longrightarrow\quad
    |X|=p.
    \label{eq:app-p-family}
\end{equation}
Write \(\mathsf{Term}(v,u,\ell)\) for a completed active component: either an
altruist-initiated path satisfying
\[
    v\in B,\qquad
    u\notin B,\qquad
    \ell\le\ell_p,
\]
or a cycle satisfying
\[
    v=u\notin B,\qquad
    \ell\le\ell_c.
\]
For a family \(\mathcal A\) and a fixed set \(S\), write
\begin{equation}
    \mathcal A\uplus S
      :=
    \{X\cup S:X\in\mathcal A,\ X\cap S=\varnothing\}.
    \label{eq:app-disjoint-insertion}
\end{equation}

\paragraph{Initialization.}
For a legal base state
\[
    k=\ell=1,\qquad
    p=2,\qquad
    (v,u)\in A,\qquad
    v\ne u,
\]
the active component is the single exchange edge \(v\to u\), and
\begin{equation}
    \mathcal F_{1,1,2}^{vu}
      =
    \{\{v,u\}\}.
    \label{eq:app-tx-base}
\end{equation}

\paragraph{Starting a new component.}
For \(k>1\), \(\ell=1\), and a legal new active edge
\((v,u)\in A\), a new component can start only after the preceding active
component is terminal:
\begin{align}
\mathcal F_{k,1,p}^{vu}
={}&
 \bigcup_{\substack{v'\in B,\ u'\notin B\\1\le\ell'\le\ell_p}}
 \left(
    \mathcal F_{k-1,\ell',p-2}^{v'u'}
    \uplus\{v,u\}
 \right)
 \nonumber\\[-1mm]
&\quad\cup
 \bigcup_{\substack{v'\notin B\\1\le\ell'\le\ell_c}}
 \left(
    \mathcal F_{k-1,\ell',p-2}^{v'v'}
    \uplus\{v,u\}
 \right),
\label{eq:app-tx-start}
\end{align}
where only semantically valid predecessor states contribute nonempty
families.  The disjoint insertion of \(\{v,u\}\) ensures that both vertices
of the new active edge are unused, and the used-set cardinality changes from
\(p-2\) to \(p\).

\paragraph{Extending the active component.}
For every semantically valid path-extension state,
\begin{equation}
    \mathcal F_{k,\ell,p}^{vu}
      =
    \bigcup_{\substack{w\ne v\\(w,u)\in A}}
    \left(
       \mathcal F_{k-1,\ell-1,p-1}^{vw}
       \uplus\{u\}
    \right).
    \label{eq:app-tx-extend}
\end{equation}
The newly reached vertex \(u\) must be unused, and the used-set cardinality
changes from \(p-1\) to \(p\).

\paragraph{Closing a cycle.}
For a legal closing state \(v=u\notin B\) with
\(1<\ell\le\ell_c\),
\begin{equation}
    \mathcal F_{k,\ell,p}^{vv}
      =
    \bigcup_{\substack{w\ne v\\(w,v)\in A}}
    \mathcal F_{k-1,\ell-1,p}^{vw}.
    \label{eq:app-tx-close}
\end{equation}
Here, the closing arc is \(w\to v\).  The vertex \(v\) is already present in
the predecessor used set, so closing the cycle introduces no new graph
vertex and the used-set cardinality remains \(p\).

\paragraph{DRS-relevant algebra.}
Every non-base predecessor lies at exchange length \(k-1\).  Each recurrence
term is obtained from a predecessor family by disjointly inserting at most
two fixed vertices, and alternative terms are combined by Boolean union.
Specifically, starting a new component inserts two vertices, extending the
active component inserts one, and closing a cycle uses union only.  The
recurrence never deletes a used vertex, joins two independently evolving
used sets, or performs arbitrary subset convolution.  These are exactly the
structural properties used by the DRS invariant in
Appendix~\ref{app:drs-details}.

%% file: appendices/C_drs_invariant_details.tex
\section{Detailed Proof of the Statewise Extension Invariant}
\label{app:drs-details}

We give the case-by-case proof of the statewise DRS invariant.  Throughout,
the semantic families are those of Appendix~\ref{app:tx-reconstruction},
and all graph-theoretic legality conditions are evaluated by the outer
dynamic program before the corresponding DRS operations are invoked.

Fix the global DRS budget
\begin{equation}
    K=2t.
    \label{eq:app-global-budget}
\end{equation}
For a state \(\sigma=(k,\ell,p,v,u)\), every
\(X\in\mathcal F_\sigma\) has \(|X|=p\).  An admissible future set satisfies
\(|Y|\le K-p\), and therefore
\begin{equation}
    |X|=p\le K-|Y|.
    \label{eq:app-query-cardinality}
\end{equation}
Thus, the cardinality side condition in the formal DRS query is automatic;
it remains only to preserve disjointness.

\paragraph{Initialization.}
For a legal base state, the semantic family is
\[
    \mathcal F_{1,1,2}^{vu}=\{\{v,u\}\}.
\]
Starting from the DRS object representing \(\{\varnothing\}\) and applying
element convolution with \(v\) and then \(u\) yields, under the global budget
\(K=2t\), a DRS object query-equivalent to this singleton family.  Since the
graph is loopless, \(v\ne u\), so the resulting used set has cardinality
exactly two.

\paragraph{General insertion identity.}
Suppose a transition inserts a fixed set \(S\) of \(s\) new vertices and
takes a predecessor of cardinality \(p-s\) to a successor of cardinality
\(p\).  For every predecessor set \(X\) and future set \(Y\),
\begin{equation}
    (X\cup S)\cap Y=\varnothing
    \quad\Longleftrightarrow\quad
    S\cap Y=\varnothing
    \ \text{and}\ 
    X\cap(Y\cup S)=\varnothing.
    \label{eq:app-insertion-identity}
\end{equation}
If \(S\cap Y\ne\varnothing\), this transition term contributes no successor
disjoint from \(Y\).  Otherwise, for \(|Y|\le K-p\),
\begin{equation}
    |Y\cup S|
      =
    |Y|+s
      \le
    K-p+s
      =
    K-(p-s).
    \label{eq:app-residual-budget}
\end{equation}
Hence, the induced predecessor query lies within exactly the residual budget
of a \((p-s)\)-vertex state.  Equation~\eqref{eq:app-residual-budget} is the
local budget identity used in every recurrence case.

\paragraph{Two inserted vertices.}
For a new-component transition, \(S=\{v,u\}\) and the predecessor has
cardinality \(p-2\).  If \(\{v,u\}\cap Y=\varnothing\), then
\[
    (X\cup\{v,u\})\cap Y=\varnothing
    \quad\Longleftrightarrow\quad
    X\cap(Y\cup\{v,u\})=\varnothing,
\]
and
\[
    |Y\cup\{v,u\}|
      \le
    2t-p+2
      =
    2t-(p-2).
\]
Extension equivalence of the predecessor therefore gives the correct answer
for this transition term.  Two-element convolutions implement the
corresponding disjoint insertion while preserving the required query
semantics, and Boolean DRS union combines the legal terminal predecessors.

\paragraph{One inserted vertex.}
For an extension transition, \(S=\{u\}\) and the predecessor has cardinality
\(p-1\).  If \(u\notin Y\), then
\[
    (X\cup\{u\})\cap Y=\varnothing
    \quad\Longleftrightarrow\quad
    X\cap(Y\cup\{u\})=\varnothing,
\]
with
\[
    |Y\cup\{u\}|
      \le
    2t-p+1
      =
    2t-(p-1).
\]
Thus, one element convolution implements the required disjoint insertion and
preserves exactly the successor queries.  If \(u\in Y\), this transition
term contributes no positive answer.

\paragraph{No inserted vertex and union.}
A cycle-closing transition has \(S=\varnothing\) and leaves cardinality \(p\)
unchanged, so it passes the same future set \(Y\) to its predecessor.  For a
union
\[
    \mathcal F_\sigma=\bigcup_{i\in I}\mathcal A_i,
\]
we have
\begin{equation}
    \exists X\in\mathcal F_\sigma:X\cap Y=\varnothing
    \quad\Longleftrightarrow\quad
    \exists i\in I\ \exists X\in\mathcal A_i:X\cap Y=\varnothing.
    \label{eq:app-union-query}
\end{equation}
Hence, the Boolean DRS union preserves the existential query exactly.

\paragraph{Induction over exchange length.}
Every non-base transition has predecessor exchange length \(k-1\).
Initialization gives the base case, and the cases above give the induction
step.  Consequently, for every maintained state
\(\sigma=(k,\ell,p,v,u)\) with \(p\le2t\) and every
\(Y\subseteq V\) with \(|Y|\le2t-p\),
\begin{equation}
    \Query(\mathfrak R_\sigma,Y)=1
    \quad\Longleftrightarrow\quad
    \exists X\in\mathcal F_\sigma:X\cap Y=\varnothing.
    \label{eq:app-statewise-invariant}
\end{equation}
In particular, taking \(Y=\varnothing\) tests precisely whether the
corresponding exact semantic family is nonempty.

\paragraph{Why one global budget suffices on an accepting derivation.}
Tian and Xiao's bounded-component analysis supplies a relevant accepting
terminal state with
\[
    t\le k\le p\le2t
\]
\cite{TianXiao2026}.  Tracing a constructive recurrence derivation backward
from such a terminal state preserves the same bound: every recurrence step
only inserts used vertices and never deletes them, so each predecessor used
set is a subset of the terminal used set. Hence, every state on this
accepting derivation has cardinality at most \(2t\) and is covered by the
single global choice \(K=2t\).

%% file: appendices/D_weighted_details.tex
\section{Weighted Exact-Cardinality Extension}
\label{app:weighted-details}

We expand the min-plus argument for Exact-\(t\) Weighted Kidney Exchange.
Let \(w(e)\in\mathbb Z_{\ge0}\) be word-size arc utilities, set
\begin{equation}
    M:=\max\bigl(\{0\}\cup\{w(e):e\in A\}\bigr),
    \qquad
    d(e):=M-w(e)\ge0,
    \label{eq:app-weight-transform}
\end{equation}
and write
\[
    W(\mathcal D)
      :=
    \sum_{e\in E(\mathcal D)}w(e).
\]
Every exact-\(t\) packing contains exactly \(t\) transplant arcs, so
\begin{equation}
    C(\mathcal D)
      :=
    \sum_{e\in E(\mathcal D)}d(e)
      =
    tM-W(\mathcal D).
    \label{eq:app-cost-utility}
\end{equation}
Thus, at fixed transplant cardinality \(t\), minimizing transformed cost is
equivalent to maximizing utility.

\subsection{Min-plus semantic recurrence}

For each semantic state \(\sigma\) and used set \(X\), let
\(f_\sigma(X)\) be the minimum transformed cost of a semi-feasible packing
realizing exactly \(X\), and set \(f_\sigma(X)=\infty\) if no such packing
exists.  The corresponding min-plus DRS query is
\begin{equation}
    Q_{f_\sigma}(Y)
      :=
    \min\bigl\{
        f_\sigma(X):
        X\cap Y=\varnothing,\,
        |X|\le K-|Y|
    \bigr\}.
    \label{eq:app-weighted-query}
\end{equation}
Nederlof's dynamic-representative framework admits this min-plus
specialization \cite{Nederlof2026}.  Alternative predecessor terms are
combined by minimum.  The base edge \((v,u)\) contributes \(d(v,u)\);
starting a new component with edge \((v,u)\) contributes \(d(v,u)\);
extending the active component through \((w,u)\) contributes \(d(w,u)\);
and closing a cycle through \((w,v)\) contributes \(d(w,v)\).
Consequently, every transplant arc is charged exactly once.

The disjointness argument is identical to the Boolean case.  If a
transition inserts a fixed set \(S\), with \(|S|=s\), and takes a
\((p-s)\)-vertex predecessor to a \(p\)-vertex successor, then for every
\(|Y|\le K-p\) with \(S\cap Y=\varnothing\),
\begin{equation}
    (X\cup S)\cap Y=\varnothing
      \quad\Longleftrightarrow\quad
    X\cap(Y\cup S)=\varnothing,
    \qquad
    |Y\cup S|\le K-(p-s).
    \label{eq:app-weighted-budget}
\end{equation}
The added scalar transition cost is independent of the future set \(Y\).
Hence, the same residual-budget induction applies in the min-plus semiring,
and the weighted recurrence is maintained under the global budget
\begin{equation}
    K=2t.
    \label{eq:app-weighted-global-budget}
\end{equation}

\subsection{Solutions containing one component of length exactly \(t\)}

An exact-\(t\) packing containing a component of exchange length \(t\)
has no other nonempty component.  We optimize the two possible component
types separately.

To force the auxiliary source/sink arcs while retaining the transformed
cost as a secondary objective, define
\begin{equation}
    H:=tM+1.
    \label{eq:app-forcing-offset}
\end{equation}
On every original arc \(e\), use shifted cost
\begin{equation}
    \widehat d(e):=H+d(e),
    \label{eq:app-shifted-cost}
\end{equation}
while every auxiliary gadget arc has cost zero.  Since \(0\le d(e)\le M\),
the offset \(H\) dominates the total transformed-cost variation among at
most \(t\) original arcs.

\paragraph{Exact-\(t\) altruist-initiated chain.}
Add a fresh source \(s\) and a zero-cost arc \(s\to b\) for every altruist
\(b\in B\).  Give every original arc \(e\) shifted cost
\(\widehat d(e)=H+d(e)\), and compute a minimum-cost simple directed path on
exactly \(t+2\) vertices.

Any such path that contains \(s\) must start at \(s\), because \(s\) has no
incoming arcs.  It then uses one zero-cost arc \(s\to b\) and exactly \(t\)
original arcs, and therefore corresponds bijectively to an
altruist-initiated chain with exactly \(t\) transplants.  Its shifted cost is
\begin{equation}
    tH+C(\mathcal D)
    \le tH+tM.
    \label{eq:app-chain-desired-cost}
\end{equation}
A \(t+2\)-vertex path avoiding \(s\) uses \(t+1\) original arcs and hence
has shifted cost at least
\[
    (t+1)H.
\]
Since \(H=tM+1\),
\[
    tH+tM<(t+1)H.
\]
Therefore, whenever an exact-\(t\) altruistic chain exists, every global
minimum below \((t+1)H\) uses the source gadget, and subtracting \(tH\)
recovers its transformed cost.  Conversely, an optimum of at least
\((t+1)H\) certifies that no such chain exists.

Nederlof's Weighted Directed Path algorithm thus finds the best exact-\(t\)
chain in
\begin{equation}
    2^{t+O(\sqrt t\log^2t)}n^{O(1)}
    \label{eq:app-weighted-chain-runtime}
\end{equation}
time; the additive constant in the path parameter \(t+2\) is absorbed by
the asymptotic bound.

\paragraph{Exact-\(t\) cycle.}
Fix a candidate closing arc \((u,v)\in A\).  Add fresh vertices \(s,z\),
the zero-cost arcs
\[
    s\to v,
    \qquad
    u\to z,
\]
and give every original arc shifted cost \(H+d(e)\).
Compute a minimum-cost simple directed path on exactly \(t+2\) vertices.

A path using both zero-cost gadget arcs necessarily has the form
\[
    s\to v\rightsquigarrow u\to z,
\]
where the middle segment is a simple directed \(v\to u\) path on exactly
\(t\) original vertices and hence contains exactly \(t-1\) original arcs.
Its shifted path cost is at most
\begin{equation}
    (t-1)H+(t-1)M.
    \label{eq:app-cycle-desired-cost}
\end{equation}
Any \(t+2\)-vertex path using at most one of the two gadget arcs contains at
least \(t\) original arcs and therefore has shifted cost at least \(tH\).
Because
\[
    (t-1)H+(t-1)M<tH
\]
for \(H=tM+1\), an optimum below \(tH\) necessarily uses both gadget arcs.
After subtracting \((t-1)H\), its residual cost is exactly the transformed
cost of the \(v\to u\) segment.  Adding the closing cost \(d(u,v)\) gives
the transformed cost of the corresponding exact-\(t\) cycle.

Enumerating the closing arc and taking the minimum therefore yields the
best exact-\(t\) cycle in
\begin{equation}
    2^{t+O(\sqrt t\log^2t)}n^{O(1)}
    \label{eq:app-weighted-cycle-runtime}
\end{equation}
time.

\subsection{All components shorter than \(t\)}

It remains to consider exact-\(t\) packings in which every component has
exchange length at most \(t-1\).  We therefore replace the original caps by
\begin{equation}
    \ell_p':=\min\{\ell_p,t-1\},
    \qquad
    \ell_c':=\min\{\ell_c,t-1\},
    \label{eq:app-weighted-caps}
\end{equation}
and run the min-plus semi-feasible recurrence up to total exchange length
\(k=t\).

If an exact-\(t\) packing contains \(r\) nonempty altruist-initiated chains,
then its used-vertex count is
\begin{equation}
    p=t+r\le2t.
    \label{eq:app-weighted-used-count}
\end{equation}
Indeed, every cycle contributes one used vertex per transplant arc, whereas
each nonempty altruist-initiated chain contributes one additional altruist.
Since every nonempty chain contains at least one transplant arc, \(r\le t\).

Every predecessor on a constructive derivation uses a subset of the terminal
used set. Hence, all states on an accepting exact-\(t\) derivation satisfy
\(p\le2t\), and the min-plus DRS recurrence applies throughout under the
single global budget \(K=2t\).  Its running time is
\begin{equation}
    2^{2t+O(\sqrt t\log^2t)}n^{O(1)}.
    \label{eq:app-weighted-drs-runtime}
\end{equation}

Taking the minimum transformed cost over the exact-\(t\) altruistic-chain
case, the exact-\(t\) cycle case, and the bounded-component case, and then
using \eqref{eq:app-cost-utility}, yields the maximum additive utility among
all feasible packings with exactly \(t\) transplants.

The statement is deliberately exact-cardinality: no truncation argument is
used to claim preservation of utility for arbitrary feasible packings with
more than \(t\) transplants.

%% file: references.bib
@inproceedings{MaitiDey2022,
  author    = {Arnab Maiti and Palash Dey},
  title     = {Parameterized Algorithms for Kidney Exchange},
  booktitle = {Proceedings of the Thirty-First International Joint Conference on Artificial Intelligence (IJCAI)},
  pages     = {405--411},
  year      = {2022},
  doi       = {10.24963/ijcai.2022/58}
}

@inproceedings{HebertJohnsonEtAl2024,
  author    = {\'{U}rsula H\'{e}bert-Johnson and Daniel Lokshtanov and Chinmay Sonar and Vaishali Surianarayanan},
  title     = {Parameterized Complexity of Kidney Exchange Revisited},
  booktitle = {Proceedings of the Thirty-Third International Joint Conference on Artificial Intelligence (IJCAI)},
  pages     = {76--84},
  year      = {2024},
  doi       = {10.24963/ijcai.2024/9}
}

@article{TianXiao2026,
  author  = {Kangyi Tian and Mingyu Xiao},
  title   = {A Faster Deterministic Algorithm for Kidney Exchange via Representative Set},
  journal = {arXiv preprint arXiv:2603.18471},
  year    = {2026},
  note    = {Accepted at IJCAI 2026}
}

@inproceedings{Nederlof2026,
  author    = {Jesper Nederlof},
  title     = {Weighted $k$-Path and Other Problems in Almost $O^*(2^k)$ Deterministic Time via Dynamic Representative Sets},
  booktitle = {2025 IEEE 66th Annual Symposium on Foundations of Computer Science (FOCS)},
  pages     = {2813--2824},
  year      = {2026},
  publisher = {IEEE Computer Society},
  doi       = {10.1109/FOCS63196.2025.00143}
}

@article{Zehavi2016,
  author  = {Meirav Zehavi},
  title   = {A Randomized Algorithm for Long Directed Cycle},
  journal = {Information Processing Letters},
  volume  = {116},
  number  = {6},
  pages   = {419--422},
  year    = {2016},
  doi     = {10.1016/j.ipl.2016.02.005}
}

@article{FominEtAl2016,
  author  = {Fedor V. Fomin and Daniel Lokshtanov and Fahad Panolan and Saket Saurabh},
  title   = {Efficient Computation of Representative Families with Applications in Parameterized and Exact Algorithms},
  journal = {Journal of the ACM},
  volume  = {63},
  number  = {4},
  pages   = {29:1--29:60},
  year    = {2016},
  doi     = {10.1145/2886094}
}

@inproceedings{BanikEtAl2026,
  author    = {Aritra Banik and Sujoy Bhore and Palash Dey and Abhishek Sahu},
  title     = {Kidney Exchange: Faster Parameterized Algorithms and Tighter Lower Bounds},
  booktitle = {Proceedings of the 25th International Conference on Autonomous Agents and Multiagent Systems (AAMAS)},
  pages     = {2708--2716},
  year      = {2026},
  publisher = {IFAAMAS}
}

@article{FominEtAl2018LongPath,
  author  = {Fedor V. Fomin and Daniel Lokshtanov and Fahad Panolan and Saket Saurabh and Meirav Zehavi},
  title   = {Long Directed $(s,t)$-Path: {FPT} Algorithm},
  journal = {Information Processing Letters},
  volume  = {140},
  pages   = {8--12},
  year    = {2018},
  doi     = {10.1016/j.ipl.2018.04.018}
}

@article{GabowNie2008,
  author  = {Harold N. Gabow and Shuxin Nie},
  title   = {Finding a Long Directed Cycle},
  journal = {ACM Transactions on Algorithms},
  volume  = {4},
  number  = {1},
  pages   = {7:1--7:21},
  year    = {2008},
  doi     = {10.1145/1328911.1328918}
}

@article{LinEtAl2019,
  author  = {Mugang Lin and Jianxin Wang and Qilong Feng and Bin Fu},
  title   = {Randomized Parameterized Algorithms for the Kidney Exchange Problem},
  journal = {Algorithms},
  volume  = {12},
  number  = {2},
  pages   = {50},
  year    = {2019},
  doi     = {10.3390/a12020050}
}

@article{ManloveOMalley2014,
  author  = {David F. Manlove and Gregg O'Malley},
  title   = {Paired and Altruistic Kidney Donation in the {UK}: Algorithms and Experimentation},
  journal = {ACM Journal of Experimental Algorithmics},
  volume  = {19},
  pages   = {2.6:1--2.6:21},
  year    = {2014},
  doi     = {10.1145/2670129}
}

@inproceedings{DickersonEtAl2016,
  author    = {John P. Dickerson and David F. Manlove and Benjamin Plaut and Tuomas Sandholm and James Trimble},
  title     = {Position-Indexed Formulations for Kidney Exchange},
  booktitle = {Proceedings of the 2016 ACM Conference on Economics and Computation},
  pages     = {25--42},
  year      = {2016},
  doi       = {10.1145/2940716.2940759}
}

@article{BarkelEtAl2026,
  author  = {Mathijs Barkel and Rachael Colley and Maxence Delorme and David Manlove and William Pettersson},
  title   = {Operational Research Approaches and Mathematical Models for Kidney Exchange: A Literature Survey and Empirical Evaluation},
  journal = {European Journal of Operational Research},
  volume  = {331},
  number  = {2},
  pages   = {325--350},
  year    = {2026},
  doi     = {10.1016/j.ejor.2025.08.059}
}

@article{BiroManloveRizzi2009,
  author  = {P{\'e}ter Bir{\'o} and David F. Manlove and Romeo Rizzi},
  title   = {Maximum Weight Cycle Packing in Directed Graphs, with Application to Kidney Exchange Programs},
  journal = {Discrete Mathematics, Algorithms and Applications},
  volume  = {1},
  number  = {4},
  pages   = {499--517},
  year    = {2009},
  doi     = {10.1142/S1793830909000373}
}

@article{DelormeEtAl2024,
  author  = {Maxence Delorme and Sergio Garc{\'i}a and Jacek Gondzio and J{\"o}rg Kalcsics and David Manlove and William Pettersson},
  title   = {New Algorithms for Hierarchical Optimization in Kidney Exchange Programs},
  journal = {Operations Research},
  volume  = {72},
  number  = {4},
  pages   = {1654--1673},
  year    = {2024},
  doi     = {10.1287/opre.2022.2374}
}
